\documentclass[11pt]{article}
\usepackage{amsmath,amsfonts,amssymb,amsthm,bbm,bm}
\usepackage{pdfsync}
\usepackage{graphicx}
\usepackage[latin1]{inputenc}
\usepackage{hyperref}
\usepackage[all]{xy}
\usepackage{stmaryrd}
\usepackage{rotating}
\usepackage{color}

\newcommand{\bit}{\begin{itemize}}
\newcommand{\eit}{\end{itemize}}
\newcommand{\bd}{\begin{description}}
\newcommand{\ed}{\end{description}}

\newcommand{\bc}{\begin{center}}
\newcommand{\ec}{\end{center}}
\newcommand{\Ref}[1]{(\ref{#1})}

\newcommand{\SO}{\mathrm{SO}}

\newcommand{\scr}{\scriptscriptstyle\rm}

\newcommand{\be}{\begin{equation}}
\newcommand{\ee}{\end{equation}}
\newcommand{\bea}{\begin{eqnarray}}
\newcommand{\eea}{\end{eqnarray}}
\newcommand{\bs}{\begin{subequations}}
\newcommand{\es}{\end{subequations}}
\newcommand{\nn}{\nonumber}

\newcommand{\w}{\wedge}

\newcommand{\f}{\frac}
\newcommand{\tl}{\tilde}
\def\p{\partial}

\newcommand{\na}{\nabla}

\def\a{\alpha}
\def\b{\beta}
\def\g{\gamma}
\def\d{\delta}
\def\eps{\epsilon}

\def\th{\theta}

\def\k{\kappa}
\def\l{\lambda}
\def\m{\mu}
\def\n{\nu}

\def\r{\rho}

\def\s{\sigma}

\def\om{\omega}
\def\G{\Gamma}

\def\Th{\Theta}

\def\Si{\Sigma}

\def\L{\Lambda}
\def\Om{\Omega}

\newcommand{\scri}{\cal I}

\newcommand{\os}[1]{\overset{\circ}{#1}}
\newcommand{\og}[1]{\overset{\scriptscriptstyle e}{#1}{}}
\newcommand{\ut}[1]{ \underset{\widetilde{}}{#1}{} }

\newcommand{\dt}{{-\hspace{-6.4pt}\d}}

\begin{document}
%%%%%%%%%%%%%%%%%%%%%%%%%%%%%%%%%%%%%%%%%%%%%%%%%%%%%%%%

\title{\bf A gauge-invariant symplectic potential for tetrad general relativity } %Surface charges in real connection variables}

\author{\Large{Elena De Paoli$^{1,2}$ and Simone Speziale$^1$}
\smallskip \\ \small{$^1$ Aix Marseille Univ., Univ. de Toulon, CNRS, CPT, UMR 7332, 13288 Marseille, France} \\
\small{$^2$Dip. di Fisica, Univ. di Roma 3, Via della Vasca Navale 84, 00146 Roma, Italy} 
}
\date{\today}

\maketitle

%----------------------------------------------------------------------------
\begin{abstract}
\noindent 
We identify a symplectic potential for general relativity in tetrad and connection variables that is fully gauge-invariant, using the freedom to add surface terms. When torsion vanishes, it does not lead to surface charges associated with the internal Lorentz transformations, and reduces exactly to the symplectic potential given by the Einstein-Hilbert action. In particular, it reproduces the Komar form when the variation is a Lie derivative, and the geometric expression in terms of extrinsic curvature and 2d corner data for a general variation.
The additional surface term vanishes at spatial infinity for asymptotically flat spacetimes, thus the usual Poincar\'e charges are obtained. 
We prove that the first law of black hole mechanics follows from the Noether identity associated with the covariant Lie derivative, and that it is independent of the ambiguities in the symplectic potential provided one takes into account the presence of non-trivial Lorentz charges that these ambiguities can introduce.
\end{abstract}
%----------------------------------------------------------------------------

\tableofcontents

%----------------------------------------------------------------------------
\section{Introduction}
%----------------------------------------------------------------------------

Covariant phase space methods \cite{AshtekarReula,WittenCovPhase,LeeWald,Wald:1999wa} provide powerful tools for the study of symmetries and conservation laws in gauge theories and gravity. 
In gravity most literature is based on the metric formalism, but tetrads and Lorentz connection as variables have many applications, from the description of radiative data on $\scri^+$ and of isolated and dynamical horizons, to fermion coupling and quantum gravity. 
Covariant phase space methods have been successfully applied to tetrad gravity, recovering the metric Poincar\`e charges at spatial infinity, the first law of black hole mechanics and its generalization to isolated horizons
\cite{Ashtekar:2000hw,Ashtekar:2008jw,Corichi:2013zza,Corichi:2016zac}.

Notwithstanding these positive results, the symplectic potential most commonly used has two unappealing features that we wish to improve upon, and which motivate this paper.
The first issue is that it is not fully gauge-invariant: the associated pre-symplectic form has degenerate gauge directions inside the Cauchy hypersurface, but not on its boundary, unless this is taken at infinity and with appropriate fall-off conditions. This means that the covariant phase space gives in general non-trivial surface charges for internal Lorentz transformations, even when torsion vanishes. Since when torsion vanishes we would like to recover the same physics as in the metric theory, such charges appear unphysical to us.
The second and related issue is that, again when torsion vanishes, it is not equivalent to the symplectic potential taken from the Einstein-Hilbert action. This difference shows up for instance if we look at a variation given by a Lie derivative: the familiar Komar term which appears in the metric case is not present. 
As a consequence, also the Noether charge is different, which led the authors of \cite{TedMohd,Prabhu:2015vua} to point out a potential problem with the derivation of the first law from the Noether identity, and to propose that in tetrad gravity the Noether charge for diffeomorphisms should be associated not to a Lie derivative, but to a modified derivative involving an internal gauge transformation which depends non-linearly on the tetrad. 
As we will see this is not necessary: solving the first issue automatically solves the second.

To find a fully gauge-invariant symplectic potential, it is enough to use the fact that the symplectic potential is defined from an action principle only up to the addition of an exact form. Our first result is to identify an appropriate exact form that makes the pre-symplectic form completely gauge-invariant, thus free of internal Lorentz charges in the absence of torsion.
Our second result is to show that the gauge-invariant potential gives exactly the Komar term when the variation is a Lie derivative, thus recovering the expected Noether charge.
Finally we prove equivalence to the Einstein-Hilbert symplectic potential for a general variation, in the absence of torsion, by reproducing the geometric formula of \cite{BurnettWald1990,Lehner:2016vdi} in terms of extrinsic geometry and 2d corner terms. 
Support for the gauge-invariant symplectic potential we propose comes also from the fact that it turns out to match the boundary term found in \cite{Bodendorfer:2013jba} using Hamiltonian methods, starting from the requirement of finding a canonical transformation from the tetrad to the ADM phase space in the presence of 2d corner terms. 
The importance of working with a gauge-invariant potential for generic gauge theories has been discussed in details in \cite{Barnich:2007bf}, and our construction shows how this can be done for tetrad gravity.

Having established these results, we look at physical applications, in particular to asymptotic charges and to the first law of black hole mechanics.
Since the modification we propose changes the symplectic form and thus the phase space structure, it is not guaranteed that the results in the literature still apply: the exact form affects the Hamiltonian charges of the theory. 
For asymptotic Poincar\'e charges, it is easy to see that the result of \cite{Ashtekar:2008jw} is preserved, since with those asymptotic fall-off conditions  the additional exact form vanishes at spatial infinity.

For the first law, the situation is more interesting. First of all, having recovered the equivalence with the Einstein-Hilbert symplectic potential, we can immediately show that using our gauge-invariant symplectic potential the first law follows from the Noether identity associated with covariant Lie derivatives, coherently with the metric case. However, we show that the first law follows \emph{also} from the non-gauge-invariant potential and the same Lie derivative, 
provided one takes into account the non-trivial internal Lorentz charge. The latter has the effect of changing the Hamiltonian Killing flow, because tetrads and connections are preserved by a Killing Lie derivative only up to an internal transformation, and not identically.
Recovering the first law from the non-gauge invariant potential and the Lie derivative is in fact not new: it was already proven in
\cite{Ashtekar:2000hw} 
using directly the Hamiltonian generators, not expressing them in terms of the Noether charge and thus without puzzling over that mismatch. The presence of a non-zero Hamiltonian diffeomorphism generator was indeed observed in \cite{Ashtekar:2000hw}, and referred to as the horizon energy. 
Our construction clarifies that this horizon energy is the internal Lorentz charge produced by using a non-gauge covariant potential. 

Therefore, we have a situation similar to the metric case, albeit slightly subtler. In the metric case, the first law is invariant under the cohomology ambiguity in the symplectic potential, because the contribution of the exact form to the symplectic form vanishes. In the tetrad formalism this is not the case, because the Killing Lie flow vanishes only up to internal transformations. Nonetheless, the first law is still invariant, provided one takes into account the non-trivial internal Lorentz charges that can be present changing the symplectic potential by an exact form.

We hope that our results contribute to the standing of the covariant phase space for tetrad general relativity, and clarify some aspects of the existing literature. 
In the conclusions we briefly discuss future developments and applications.
We use signature with mostly plus, greek letters for spacetime indices and capital latin letters for internal indices.

%----------------------------------------------------------------------------
\section{A gauge-invariant symplectic potential }
%----------------------------------------------------------------------------
We consider the following first order action for Einstein-Cartan gravity (for a review, see \cite{Hehl:1994ue})
\be\label{SEC}
S_{\scr EC}(e,\om) = 
P_{IJKL} \int_M e^I\w e^J\w F^{KL}(\om) - \f\L6 e^I\w e^J\w e^K\w e^L,
\ee
where we have taken units $16\pi G=1$, and
\be
P_{IJKL} := \f1{2\g}(\eta_{IK}\eta_{JL}-\eta_{IL}\eta_{JK}) +\f12\eps_{IJKL}.
\ee
The action is to be supplemented by appropriate boundary integrals $I_{3d}$ and $I_{2d}$ depending on the boundary conditions chosen, see e.g. \cite{Obukhov,Bodendorfer:2013hla,Corichi:2016zac} for 3d boundaries without corners, and \cite{SorkinCorner17} in the presence of corners. 
The coupling constant $\g$ is referred to as Barbero-Immirzi parameter in most literature,\footnote{Because of the role it plays in the canonical transformation to real Ashtekar-Barbero variables, see e.g. \cite{ThiemannBook}.} and the associated Lagrangian density corresponds to the additional dimension-two term $\tl\eps^{\m\n\r\s}R_{\m\n\r\s}(\G)$ that one can write in the first order formalism.\footnote{For the interested reader, this parameter has an interesting renormalization flow \cite{Daum:2010qt,IoDario}, with an on-shell logarithmic divergence induced by the simultaneous presence of fermions and $\Lambda$ \cite{IoDarioProc}.}
The variation of the action gives the field equations and a boundary term,
\be
\int_M d \Big(P_{IJKL} e^I\w e^J\w \d\om^{KL}\Big), 
\ee
which will be the centre of attention of this paper, for the role it plays in the covariant phase space formalism.
Let us denote by $d\th_{\scr EC}(\d)$ the integrand.

The theory defined by \Ref{SEC} differs a priori from general relativity: it is only defined for orientable manifolds, and odd under orientation inversion, instead of even; it allows for degenerate tetrads hence degenerate metrics; it allows for non-vanishing spacetime torsion $T^I:=d_\om e^I$, if matter couples to the affine connection $\om^{IJ}$. In the following, we restrict attention to an invertible, right-handed tetrad. 
Then when torsion vanishes $\om^{IJ}=\og{\om}^{IJ}$ is the Levi-Civita spin connection, \Ref{SEC} is equivalent to the Einstein-Hilbert action $S_{\scr EH}$ and thus the theory to general relativity. This equivalence extends to the boundary term:  
\be\label{dthEquiv}
d\th_{\scr EC}\big|_{\om=\og{\om}} = d\th_{\scr EH},
\ee
as can be easily seen for instance from 
\be
e^\m_I e^\n_J 2\og{D}_{[\m}\d \og{\om}^{IJ}_{\n]} = e^\m_I e^\n_J \d F^{IJ}_{\m\n}(e) = \d R - \d (e^\m_I e^\n_J) \d F^{IJ}_{\m\n}(e)
=  \d R - 2 R^I{}_\m\d e^\m_I  = g^{\m\n}\d R_{\m\n}.
\ee

The equivalence \Ref{dthEquiv} implies also that the 3-forms are equal up to an exact form,
\be\label{thda}
\th_{\scr EH} = \th_{\scr EC}|_{\om=\og{\om}} + d\a.
\ee
The question we address in this paper is to find an $\a$ for which the equality above holds. It is motivated by the covariant phase space formalism, which uses the boundary term to define Noether and Hamiltonian charges of the theory.
Let us briefly review the basic points of this formalism, referring the reader to e.g. \cite{Wald:1999wa} for details.

Suppose that the boundary $\p M$ of  $M$ (which can be the whole spacetime or just a region of interest) admits a canonical split with the identification of a Cauchy hypersurface $\Si$. Then the boundary term $d\th(\d)$ obtained from the variation of a Lagrangian 4-form $L$ can be used to provide 
a symplectic potential on the space of solutions to the field equations, by taking its integral on $\Si$:
\be
\Th(\d):= \int_\Si\th(\d).
\ee
This defines a one-form in field space, and its exterior derivative is the pre-symplectic two-form
\be\label{Omdef}
\Om(\d_1,\d_2) = \d_1\Th(\d_2) - \d_2 \Th(\d_1) - \Th([\d_1,\d_2]).
\ee
Using $\d L\approx d\th(\d)$, where here and in the following $\approx$ means on-shell of the field equations, one sees that $\Om$ is independent of the choice of hypersurface $\Si$ if the background fields as well as the variations $\d_1$ and $\d_2$ satisfy the field equations. 

The symplectic structure so defined is not unique. First, the explicit form of the potential depends also on the boundary terms $I_{3d}$ and $I_{2d}$ in the action principle. These however do not affect the pre-symplectic form since the symplectic potential is changed by a total variation, therefore the covariant phase space structure is independent of them. There is nonetheless a certain freedom, since 
the symplectic potential is defined by the Lagrangian $L$ only up to an exact form, that is the Lagrangian identifies an equivalence class
\be\label{ThAmbi}
L \ \longrightarrow \ \{\th(\d) = \th(\d) +d\a(\d)\},
\ee
where $\a$ is an arbitrary 2-form in spacetime and 1-form in field space. This cohomology freedom does affect the pre-symplectic form, and it is important to test that physical predictions are independent of it. 
This freedom plays an important role below.

The simplest set-up for this formalism is when $\p M=\Si_1\cup\Si_2$ joined at a 2d space-like surface, in which case the canonical splitting is obvious. A more general configuration is a topological cylinder, 
$\p M=\Si_1\cup\Si_2\cup {\cal T}$, with the time-like hypersurface $\cal T$ connecting the 2d space-like boundary $\p\Si_1$ to $\p\Si_2$. To introduce a canonical split in this case we typically require that $\Th(\d)$ vanishes on $\cal T$.\footnote{This is because data can generically both inflow and outflow off a time-like boundary, making a canonical split impossible without restricting the phase space. Another useful set-up is when the time-like boundary is replaced by a null hypersurface $\cal N$. In that case we can have a canonical split with non-zero contribution from $\cal N$, since it is a one-way only membrane, see e.g. \cite{Ashtekar:2000hw,Wieland:2017zkf,AbhayWolfgang}.} This is a restriction on the admissible solutions if $\cal T$ is in the spacetime bulk, but can become negligible if the boundary is pushed to infinity, and it is the fall-off conditions on the fields that guarantee the vanishing of $\Th(\d)$ on ${\cal T}_\infty$. This set-up is relevant for instance in the study of asymptotic charges at spatial infinity 
with $\L=0$. 
The appropriate fall-off conditions for \Ref{SEC} where given in \cite{Ashtekar:2008jw}. We will come back to this point below in Section~\ref{SecPoincare}. 

The power of this formalism for diff-invariant theories is that it allows one to define quasi-local Hamiltonian charges for diffeomorphisms as the canonical generators in the covariant phase space.\footnote{Let us remind the reader less familiar with this formalism that for a general diffeomorphisms, these quasi-local charges are not interesting observables, because their value depends on the shape of the boundary of $\Si$. It is only when $\xi^\m$ is a Killing or asymptotic Killing vector that the charges are truly useful.} They are given by the pre-symplectic form with one variation being a 
Lie derivative $\d_\xi=\pounds_\xi$,
\be \label{defHxi}
\dt H_{\xi}[\Si] := \Om(\d,\d_\xi) = \int_\Si \d \th(\d_\xi)-\d_\xi\th(\d).
\ee
Here we assumed that $[\d_\xi,\d]=0$,\footnote{This is a customary assumption \cite{Wald:1999wa}, although it can be argued \cite{AshtekarReula} that it is rather a \emph{definition} of what we mean by the perturbation of a diffeomorphed solution.} and the $\dt$ is there to remind us that the quantity on the RHS is not always a total variation.
Only when it is, the generator integrates to a proper Hamiltonian charge $H_\xi[\Si]$.\footnote{Since a typical case study is when $\Si$ has two boundaries, $H_\xi[\Si]$ is also referred to as a flux, leaving the name charge for the surface integrals whose difference makes up $H_\xi[\Si]$, see below.}
The integrability condition is  $\int_\Si \om(\d_1,\d_2)\lrcorner\xi=0$ \cite{Wald:1999wa}, where $\om$ is the integrand of $\Om$, and a sufficient condition familiar from the ADM energy calculations is the existence of a functional $B$ such that $\th(\d)\lrcorner\xi=\d B\lrcorner\xi$.

The origin of this latter condition becomes clear if we recall the relation between the Hamiltonian charges and 
the Noether charges, which do not coincide for diff-invariant theories.
The conserved Noether current is given by (see e.g. \cite{Iyer:1994ys})
\be\label{defj}
j(\d_\xi) := \th(\d_\xi) - L\lrcorner\xi,
\ee
since this is the object that is closed on-shell: Using 
$\d_\xi
 L 
=d(L\lrcorner\xi)$, it is immediate to see that 
$dj(\d_\xi)\approx 0$. 
Furthermore, it is also possible to show that $j(\d_\xi)\approx d q(\xi)$ for some 2-form $q(\xi)$ \cite{Iyer:1994ys}. It follows that the Noether charge, defined as the integral of the current,\footnote{Noether charges for gravity can also be derived without using covariant phase space methods, see e.g. \cite{Barnich:2000zw}. For a derivation of Noether charges for first order tetrad gravity with these methods, see \cite{Barnich:2016rwk}.} is a boundary term:
\be
Q_{\xi}[\p\Si]:= \int_\Si j(\d_\xi)\approx 
\int_{\p \Si} q(\xi).
\ee
To find the relation between the Hamiltonian and Noether charges one takes the variation of \Ref{defj}, and replaces it  in the definition \Ref{defHxi} together with the Lie derivative variation $\d_\xi\th(\d)=\pounds_\xi\th(\d)=d\th\lrcorner\xi+d(\th\lrcorner\xi)$. This gives
\be\label{HQ}
\dt H_{\xi}[\Si] \approx \d Q_{\xi}[\p\Si] -  \int_{\p\Si}\th(\d)\lrcorner\xi.
\ee
This shows $(i)$ that the Hamiltonian as well as the Noether charge are surface charges, but in general differ by a term $\th(\d)\lrcorner\xi$; and $(ii)$ that if $\th(\d)\lrcorner\xi=\d B\lrcorner\xi$, then $\dt H_{\xi}[\Si]=\d H_{\xi}[\Si]$ is a total variation and thus integrable.
In spite of their close relation, the Hamiltonian and Noether charges have an important difference: the former changes only under the ambiguity \Ref{ThAmbi} in defining the symplectic potential, whereas the Noether current $j(\d_\xi)$ and charge $Q_{\p\Si}(\xi)$ are changed also by adding boundary terms $I$ to the action, which makes them less universal objects than the Hamiltonian charges.\footnote{There is also a third ambiguity in the definition of the Noether charge itself, since one can always add an exact 2-form to it. This ambiguity will play no role in the following.}

To make this quick review more concrete, let us recall that for the Einstein-Hilbert Lagrangian $L_{\scr EH}=(R-2\L)\eps$ (without boundary terms, for simplicity), we have
\be\label{thEH}
\th_{\scr EH}(\d) = 2g^{\r[\s}\d\G^{\m]}_{\r\s} d\Si_\m,
\ee
with $d\Si_\m$ the oriented volume element. Specializing to a diffeomorphism,
\be\label{thEHxi}
\th_{\scr EH}(\pounds_\xi) = d\k(\xi) + \star(2E\lrcorner\xi)+L_{\scr EH}\lrcorner\xi,
\ee
where $\k$ is the Komar form, in components
\be\label{Komar}
{\k}_{\m\n}(\xi):=-\eps_{\m\n\r\s}\na^\r\xi^\s,
\ee
$E\lrcorner\xi := (G_{\m\n}+\L g_{\m\n})\xi^\m dx^\n$ contains the field equations, and $\star$ is the Hodge dual on spacetime forms (see the Appendix for conventions). It follows that the Noether charge associated with diffeomorphisms by \Ref{thEH} is the Komar form,
\be
j(\d_\xi) = \th_{\scr EH}(\d_\xi) - L_{\scr EH}\lrcorner\xi  \approx d\k(\xi).
\ee
It also enters the  Hamiltonian charge, 
\begin{subequations} \begin{align}\label{defHxia}
\dt H_{\xi}[\Si] := \Om_{\scr EH}(\d,\d_\xi) &= \int_\Si \d \th_{\scr EH}(\d_\xi)-\d_\xi\th_{\scr EH}(\d) \\
&= \int_{\p\Si} \d \k(\xi)- \th_{\scr EH}(\d)\lrcorner\xi. \label{defHxib}
\end{align}\end{subequations}
This equation is the starting point to prove the first law of black hole mechanics.
 
Coming back to the tetrad action \Ref{SEC}, we see that it defines the symplectic potential\footnote{Another common choice is the opposite polarization, obtained adding the extrinsic geometry boundary term to the action. All considerations in this paper apply also to this alternative choice, although some explicit formulae are different.}
\be\label{Th0}
\Th_{\scr EC}(\d):=\int_\Si \th_{\scr EC}(\d) := \int_\Si P_{IJKL}e^I\w e^J\w \d\om^{KL}.
\ee
This turns out not to be equivalent to \Ref{thEH} when torsion vanishes, hence a non-zero $\alpha$ is required in \Ref{thda}.
The difference shows up prominently when one evaluates the symplectic potential for a diffeomorphism variation $\d_\xi$. In the metric case with the Einstein-Hilbert Lagrangian $L_{\scr EH}$, we have \Ref{thEHxi} with the Komar form.  
When using tetrads as fundamental variables, we have the additional gauge freedom of performing internal Lorentz transformations.
The action \Ref{SEC} is thus invariant under $\SO(3,1)$ gauge transformations 
\be
\d_\l e^I =\l^I{}_J e^J, \qquad \d_\l\om^{IJ} = -d_\om \l^{IJ}
\ee
as well as the usual diffeomorphisms,
\begin{subequations}\label{diffeos}\begin{align}
& \pounds_\xi e^I = de^I\lrcorner\xi+d(e^I\lrcorner\xi) = d_\om e^I\lrcorner\xi+d_\om(e^I\lrcorner\xi) - (\om^I{}_J\lrcorner\xi) e^J\\
& \pounds_\xi \om^{IJ} = d\om^{IJ}\lrcorner\xi+d(\om^{IJ}\lrcorner\xi)=F^{IJ}\lrcorner\xi+d_\om(\om^{IJ}\lrcorner\xi),
\end{align}\end{subequations}
as well as combinations of the two. In particular, we can consider the gauge-covariant diffeomorphisms
\begin{subequations}\label{covdiffeos}\begin{align}
& L_\xi e^I = d_\om e^I\lrcorner\xi+d_\om(e^I\lrcorner\xi)\\
& L_\xi \om^{IJ} = d_\om\om^{IJ}\lrcorner\xi+d_\om(\om^{IJ}\lrcorner\xi) - d(\om^{IJ}\lrcorner\xi) =F^{IJ}\lrcorner\xi,
\end{align}\end{subequations}
which are defined adding a gauge transformation to the Lie derivative,
\be\label{defL}
L_\xi := \pounds_\xi + \d_{\om\lrcorner\xi}.
\ee
These are gauge-covariant, unlike \Ref{diffeos}, and $[L_\xi,\d_\l]=\d_{d_\om\l\lrcorner\xi}$ is a gauge transformation.\footnote{The reader familiar with the Hamiltonian analysis of \Ref{SEC} will recognise these two covariances as those associated respectively to the generators 
\be\nn
{\cal D}_a := {\cal C}_a - \om^{IJ}_a {\cal G}_{IJ}, \qquad
{\cal C}_a:= -2\tl P{}^b_{IJ} F^{IJ}_{ab}. 
\ee
see e.g. \cite{ThiemannBook}.
}

Taking variations given by these Lie derivatives, the potential \Ref{Th0} gives
\begin{subequations}\label{ThAch}\begin{align}
\th_{\scr EC}(\pounds_\xi)&= P_{IJKL}\Big[ e^I\w e^J \w F^{KL}\lrcorner\xi + 2e^I\w T^J \, \om^{KL}\lrcorner\xi \Big] + d(P_{IJKL} e^I\w e^J \, \om^{KL}\lrcorner\xi),\\ 
\th_{\scr EC}(L_\xi)&= P_{IJKL} e^I\w e^J \w F^{KL}\lrcorner\xi.\label{ThECxi}
\end{align}\end{subequations}
Using 
\begin{align}
P_{IJKL} e^I\w e^J \w F^{KL}\lrcorner\xi &= \f1{3!} \Big(- \f1\g \eps^{\a\b\g\m}F_{\a\b\g\l}\xi^\l + 2e F^{\m}{}_\l \xi^\l \Big) \eps_{\m\n\r\s}dx^\n\w dx^\r\w dx^\s 
\nn \\& 
\stackrel{{\om=\og{\om}}}{=} \big((\star2E(e)+L_{\scr EH})\lrcorner\xi\big), \label{Eonshell}
\end{align}
we see that both options differ from \Ref{thEHxi}, even when torsion vanishes.
The associated torsionless Noether current is
\begin{align} \label{Ach}
j(\pounds_\xi) &|_{\om=\og{\om}} = \star(2E(e)\lrcorner\xi) + d(P_{IJKL} e^I\w e^J \, \og{\om}^{KL}\lrcorner\xi),
\end{align}
which is exact on-shell as expected, but lacks the Komar term \Ref{Komar}, as also the current associated to $L_\xi$ would. 
Hence \Ref{Th0} does not  reproduce the Noether charge of the metric theory with neither $\pounds_\xi$ nor $L_\xi$. This does not affect the evaluation of the asymptotic Poincar\'e charges, see below in Section~\ref{SecPoincare}, but it was argued in  \cite{TedMohd} to spoil the first law of black hole mechanics. 
The solution there proposed 
was to associate the diffeomorphism Noether charge not to the original Lie derivative, but to the following mixing of diffeomorphisms and gauge transformations, 
\be\label{defLieK}
K_\xi^{(e)}e^{I} := L_\xi e^{I} + \left(e^{\n[I}\pounds_\xi e^{J]}_\n\right) e_J. 
\ee
This indeed produces the Komar term (as shown in \cite{TedMohd,Prabhu:2015vua}, or by direct evaluation of \Ref{Th0} with $\d\om^{IJ}
=K_\xi^{(e)}\og{\om}^{IJ}$), and the same proposal has been followed for instance in \cite{Montesinos:2017epa,Frodden:2017qwh}. However, this is not the origin of the alleged problem with the first law, which as we show below in Section~\ref{SecFirst} can be derived also from the Noether identity with the covariant Lie derivative. 
The key point is that the symplectic potential \Ref{Th0} does not define a gauge-invariant symplectic structure.
To see this, we look at the pre-symplectic form derived from \Ref{Th0}.
Using the shorthand notation $\Si^{IJ}:=e^I\w e^J$ and the commutativity $[\d_\l,\d]=0$ of gauge transformations and variations of the fundamental fields, we have
\begin{align}
\Om_{\scr EC}(\d,\d_\l) = \d\Th_{\scr EC}(\d_\l) - \d_\l\Th_{\scr EC}(\d)
&= - P_{IJKL} \int_\Si [\l,\Si]^{IJ}\w \d\om^{KL} + \d\Si^{IJ}\w d_\om\l^{KL} = \nn \\
& = P_{IJKL} \int_\Si \d (d_\om\Si^{IJ}) \l^{KL} - P_{IJKL} \int_{\p\Si} \d \Si^{IJ} \l^{KL}, \label{Omegaded}
\end{align}
where we used
\be
\d\Si^{IJ}\w d_\om\l^{KL} = d(\d\Si^{IJ} \l^{KL}) - (d_\om\d\Si^{IJ}) \l^{KL} = d(\d\Si^{IJ} \l^{KL}) - \Big(\d (d_\om\Si^{IJ})+[\d\om,\Si]^{IJ}\Big) \l^{KL},
\ee
and
\be
 P_{IJKL} \Big([\l,\Si]^{IJ}\w \d\om^{KL} - [\d\om,\Si]^{IJ} \l^{KL} \Big) = 0
\ee
which follows from the Jacobi identity.

On-shell of the field equations, the first term in \Ref{Omegaded} vanishes (or in the presence of torsion it would cancel against the source term coming from the matter contribution to the symplectic potential), and we are left with a surface term, which gives the non-vanishing Lorentz charge
\be\label{Lorentzcharge}
\dt H_{\l}=\Om_{\scr EC}(\d,\d_\l) = - P_{IJKL} \int_{\p\Si} \d \Si^{IJ} \l^{KL}. 
\ee
This means that the symplectic structure induced by $\Om_{\scr EC}$ has degenerate gauge directions in the bulk of $\Si$, but not on its boundary. Again this fact is well-known in the literature, see e.g. \cite{Corichi:2016zac,Frodden:2017qwh}. While in a gauge theory this is a rather natural fact with a physical meaning, we find it unpalatable in this gravitational context because it would assign charges that are not there in the metric theory, making the covariant phase spaces inequivalent even in the absence of torsion.
A fully gauge-invariant symplectic structure can be easily obtained using the ambiguity \Ref{ThAmbi} in the definition of the symplectic potential. 
We find that the required exact form is
\be\label{defalpha}
\int_{\p\Si}
\a(\d):= \int_{\p\Si} \f1\g e^I\w \d e_I + \star  e^I\w \d e_I = -P_{IJKL}\int_{\p\Si} e^I\w e^J \, e^{\r K}\d e^L_\r. 
\ee
In fact, a simple calculation shows that
\be
\int_{\p\Si}
\d\a(\d_\l)-\d_\l\a(\d) = P_{IJKL}\int_{\p\Si} \d(e^I\w e^J) \l^{KL},
\ee
which cancels the surface term in \Ref{Omegaded}. The corrected potential
\be\label{RB30}
\boxed{\Th_{\scr EC}'(\d):=\Th_{\scr EC}(\d)+
\int_{\p\Si}
d\a(\d)= P_{IJKL}\int_\Si e^I\w e^J\w \d\om^{KL} + \int_{\p\Si} \f1\g e^I\w \d e_I + \star  e^I\w \d e_I}
\ee
is thus gauge-invariant.  
Notice also that it satisfies $\Th'_{\scr EC}(\d_\l)=0$ for vanishing torsion.\footnote{For a gauge transformation
\be\nn
\Th_{\scr EC}(\d_\l) = \int_\Si P_{IJKL}  d_\om(e^I\w e^J) \l^{KL} - \int_{\p\Si}P_{IJKL} e^I\w e^J \l^{KL}
\ee
is a pure boundary term when torsion vanishes, cancelled by the addition of \Ref{defalpha} since $e^{\r K}\d_\l e^L_\r=-\l^{KL}$. One could also use this argument to deduce the boundary subtraction term \Ref{defalpha}.}

As it turns out, the very same exact form allows us also to recover precisely the Komar expression from a Lie derivative variation. To see this, let us first notice the following identity 
\begin{align}
e^{\n[I}L_\xi e^{J]}_\n = D^{[I}\xi^{J]} + T^{[I}_{\m\n}e^{J]\m}\xi^\n, \label{id4}
\end{align}
where 
$D_I\xi_J = e^\m_I e_J^\n \na_\m\xi_\n$ is the covariant derivative corresponding to $d_\om$.
This implies that
\begin{align}\label{aL}
& \a(L_\xi) \big|_{\om^{IJ}=\og{\om}^{IJ}}{=} -P_{IJKL} e^I\w e^J D^K\xi^L = 
\k(\xi) -\f1{2\g}\star\!\k(\xi).
\end{align}
The last piece is the trivial Komar charge $\star\k_{\m\n}(\xi) = 2 \og{\na}_{[\m}\xi_{\n]} $, similar to the trivial charge associated with the topological Lagrangian $\eps^{\m\n\r\s} F_{\m\n} F_{\r\s}$ in YM theory. This is an exact form; it does not contribute to the boundary integral, and we disregard it in the following.\footnote{It would be however non-trivial in the presence of torsion.} 
Putting \Ref{aL} together with \Ref{ThECxi} we find
\begin{align}\label{thEqui}
& \th'_{\scr EC}(L_\xi) = \th_{\scr EC}(L_\xi) + d\a(L_\xi)
\stackrel{\om=\og{\om}}{=} P_{IJKL} e^I\w e^J \w F^{KL}\lrcorner\xi + d \k(\xi) \equiv  \th_{\scr EH}(\pounds_\xi),
\end{align}
where in the last step we used \Ref{Eonshell}.
The gauge-invariant symplectic potential \Ref{RB30} reproduces precisely the metric result in the absence of torsion.\footnote{We point out that this equality holds also with the non-gauge-covariant derivative, since 
$\th'_{\scr EC}(\pounds_\xi) =\th'_{\scr EC}(L_\xi) + 2P_{IJKL}e^I\w T^J \, \om^{KL}\lrcorner\xi $.}

For these reasons, it seems to us that \Ref{RB30}
provides a better symplectic potential for the EC theory than the simple boundary term alone: it satisfies our desiderata
\be\label{uno}
\Th_{\scr EC}'(L_\xi) \big|_{\om^{IJ}=\og{\om}^{IJ}}{=} \Th_{\scr EH}(\pounds_\xi)
\ee
and
\be\label{due}
\Om'_{\scr EC}[\d_\l,\d] = 0.
\ee
As further support for the use of \Ref{RB30}, we remark that it matches the boundary term derived in the Hamiltonian analysis of \cite{Bodendorfer:2013jba}, starting from the requirement of having a canonical transformation of connection variables to the ADM phase space in the presence of corners. Here we derived it from the requirement of full gauge-invariance of the pre-symplectic structure in the covariant phase space.\footnote{When our paper appeared on the archives, Matthias Blau showed us some unpublished notes where he had also constructed the same gauge-invariant potential and proved the property \Ref{uno} \cite{BlauUn}.}

%----------------------------------------------------------------------------
\section{Equivalence for general variations}\label{SecEquiGen}
%----------------------------------------------------------------------------

Properties \Ref{uno} and \Ref{due} were the ones we cared the most for. However, it is only a few more steps to prove that the equivalence extends to arbitrary variations. In this Section we prove that 
\be\label{daje}
\Th'_{\scr EC}(\d)\big|_{\om^{IJ}=\og{\om}^{IJ}} = \Th_{\scr EH}(\d), 
\ee
namely that $\a(\d)$ defined in \Ref{defalpha} satisfies \Ref{thda}.

In tensor language, a geometric expression for the generic variation was given in \cite{BurnettWald1990}, and more recently rederived in \cite{Lehner:2016vdi}. Using the notation from the latter paper, one has
\be\label{thEHgen}
n_\m \th^\m_{\scr EH}(\d) = 2n_\m g^{\r[\s}\d\G^{\m]}_{\r\s} = -2\d K + K_{ab}\d q^{ab} -s \na_a\dt A^a,
\ee
where the notation is as follows: $n_\m$ is the unit normal to $\Si$, with signature $s:=n^2=\pm1$ and projector  $q_{\m\n}:=g_{\m\n}-sn_\m n_\n$;
$K_{\m\n}:=q_\m^\r q_\n^\r  \na_\r n_\s$ is the extrinsic curvature of the hypersurface. The authors of \cite{Lehner:2016vdi} pick coordinates $y^a(x^\m)$, $a=1,2,3$ to parametrize $\Si$, and $t^\m_a:=\p x^\m/y^a$ define tangent vectors and the induced metric $q_{ab}=q_{\m\n}t^\m_a t^\n_b$ with determinant $q$. Finally, $t_\m^a:=q^{ab}g_{\m\n} t^\n_b$ are the inverse tangent vectors, $\na_a$ the induced Levi-Civita covariant derivative and 
$\dt A^a:=-s t^a_\m \d n^\m$ captures the variation of the normal-tangential components of $\d g^{\m\n}$. For our purposes, it is convenient to rewrite this formula in a covariant way, without using tangent vectors and hypersurface tensors. To that end, we denote by $\hat r^\m$ the unit normal to the space-like boundary $\p\Si$ within $T^*\Si$: it satisfies $\hat r^2=-s$ and $\hat r_\m n^\m=0$ (and in the case when it is time-like we take it future oriented). Since $\d t^a_\m=\dt A^a n_\m$, the second term in \Ref{thEHgen} can be rewritten immediately in covariant form,
\be
K_{ab}\d q^{ab} = K_{\m\n}\d g^{\m\n} = -2 K^\m_I \d e^I_\m.
\ee
As for the boundary term we have
\be
-s  \int_\Si D_a\dt A^a d\Si = -s\int_{\p\Si} \hat r_a t^a_\m \d n^\m dS = -s\int_{\p\Si} \hat r_\m q^\m_\n \d n^\n dS
=-s \int_{\p\Si} \hat r_\m \d n^\m dS,
\ee
where $d\Si:=\sqrt{-sq}d^3y$ and $dS$ are the induced volume elements on $\Si$ and $\p\Si$.
Hence,\footnote{It is by the way in this covariant form that the equation is presented in \cite{BurnettWald1990}.}
\begin{align}
s \Th_{\scr EH}(\d) &= \int_\Si n_\m \th^\m_{\scr EH}(\d) d\Si \nn \\\label{daje1}
&= \int_\Si \big[-2\d (K\sqrt{-sq}) + (K_{\m\n}-K q_{\m\n})\sqrt{-sq}\d q^{\m\n}\big]d^3y -s \int_{\p\Si} \hat r_\m  \d n^\m dS.
\end{align}
We will take advantage of this formula to establish the equivalence \Ref{daje}, by proving that \linebreak $\th'_{\scr EC}(\d)=\th_{\scr EC}(\d)+d\a(\d)$ equals the RHS of \Ref{daje1} for vanishing torsion and right-handed tetrads.

First of all, we need an identity which allows us to rewrite the symplectic potential with the hypersurface unit normal $n_\m$ explicitly appearing:
\be\label{idthnn}
\f12 \eps_{IJKL}\int_\Sigma e^I\w e^J\w   \d\om^{KL}  = - 2 \int_\Sigma e\, \d\om_{I,}{}^{IJ}n_J =
- s \eps_{IJKL}\int_\Sigma e^I\w e^J\w   \d\om^L{}_M n^K n^M.
\ee
To see this, we use the tetrad identity \Ref{tetId2} before and after using $n^Kn^M=s(\eta^{KM}-q^{KM})$, getting
\begin{align}
\eps_{IJKL} \eps^{\m\n\r\s}e^I_\m e^J_\n \d\om_\r^{L}{}_{M} n_\s n^Kn^M &= 2 s e \, \d\om_{I,}{}^{IJ}n_J \nn \\\nn
& = - s \eps_{IJKL} \eps^{\m\n\r\s}e^I_\m e^J_\n (\d\om_\r^{KL} + \d\om_\r^L{}_{M}q^{KM})n_\s
\\ &= - s \eps_{IJKL} \eps^{\m\n\r\s}e^I_\m e^J_\n \d\om_\r^{KL}n_\s + 2s e \, \d\om_{K,LM}q^{KM}n^L.
\end{align}
Since in the last term we can replace $q^{KM}$ with $\eta^{KM}$ we obtain
\be
\eps_{IJKL} \eps^{\m\n\r\s}e^I_\m e^J_\n \d\om_\r^{KL}n_\s = 4 e \, \d\om_{I,}{}^{JI}n_J,
\ee 
from which \Ref{idthnn} follows. Another needful identity concerns the $1/\g$ piece of $\Th_{\scr EC}$: we have
\be\label{maggica3}
\int_\Si e_I\w e_J\w\d\om^{IJ} = \int_\Si \Big(T^I\w e_J \, (e^\r_I \d e_\r^J) - e_I\w\d T^I \Big)
- \int_{\p\Si} e^I\w \d e_I,
\ee
which can be shown using $\om^{IJ}_\m = e^{\l I}\na_\m e^J_\l$ and integrating by parts.

Next, we consider the following boundary term \cite{Obukhov,Bodendorfer:2013hla,Wieland:2013ata},
\begin{align}\label{K0}
&I_\Sigma:= 2 \int_\Sigma P_{IJKL}e^I\w e^J\w n^{K} d_\om n^L 
= 2\int_\Sigma e^\m_I D_\m n^I d\Si =: 2 \int_\Sigma \os{K} d\Si 
\end{align}
which represents an `affine' version $\os{K}$ of the extrinsic curvature -- in the sense of being defined without referring to the Levi-Civita connection --, and which reduces to the extrinsic curvature $K$ if there is no torsion. 
The  equality in the middle follows using \Ref{tetId2} and $n_I D_\m n^I=0$. Notice also that the term proportional to $1/\g$ vanishes identically. 
We then compute its variation, which gives
\begin{align}
\d I_\Sigma &= \int_\Sigma \eps_{IJKL}\left[2\d e^I\w e^J\w n^{K} d_\om n^L + e^I\w e^J\w (\d n^K d_\om n^L + n^K d_\om \d n^L + \d\om^L{}_M n^K n^M) \right]\nn \\ & \nn
=\int_\Sigma \eps_{IJKL}\Big[2\d e^I\w e^J\w n^{K} d_\om n^L + e^I\w e^J\w \Big(2\d n^K d_\om n^L  - \f s2\d\om^{KL}\Big) +2 e^I\w T^J \, n^K  \d n^L \Big] \\ &\hspace{2.2cm} + d(\eps_{IJKL}e^I\w e^J n^K \d n^L ) \label{martedi3}
\end{align}
where we used \Ref{idthnn}. 
The second term vanishes since $n^I$ is unit norm, and isolating the symplectic potential \Ref{Th0} in \Ref{martedi3} we find
\begin{align}\label{maggica1}
s \Th_{\scr EC}(\d) &=\eps_{IJKL} \int_\Sigma -\d \big(e^I\w e^J\w n^{K} d_\om n^L \big) + 2 \d e^I\w e^J\w n^{K} d_\om n^L 
+2 e^I\w T^J \, n^K  \d n^L   \\ \nn &\hspace{2.2cm} + \eps_{IJKL}\int_{\p\Si}e^I\w e^J n^K \d n^L + \f{s}\g \int_\Si e_I\w e_J\w\d\om^{IJ}.
\end{align}

We now compare this expression  for $\om=\og{\om}$ and $T=0$ with \Ref{daje1}.
The first term in \Ref{maggica1} gives immediately the first term in \Ref{daje1}, thanks to \Ref{K0}.
The matching of the second terms in \Ref{maggica1} and \Ref{daje1} is also easily established: 
\begin{align}
2\eps_{IJKL} \int_\Si \d e^I\w e^J \w n^K d_\om n^L &=-2 \int_\Si 
(q^\n_I\na_\n n^\m - e^\m_I\na_\r n^\r) 
\d e^I_\m d\Si
\\\nn &= -2 \int_\Si (
\os{K}_I{}^\m
- \os{K} e^\m_I)\d e^I_\m d\Si,
\end{align}
which coincides with the second term in \Ref{daje1} when torsion vanishes.
It remains to look at the boundary term of \Ref{martedi3}, which in tensor form gives
\begin{align}\label{Dzeko}
&  \eps_{IJKL}\int_{\p\Si}e^I\w e^J n^K\d n^L  = -4\int_{\p\Si} n_{[K}\hat r_{L]} n^K\d n^L dS \\\nn
& \qquad = -2s \int_{\p\Si} \hat r_{L}\d n^L dS = -2s \int_{\p\Si} (\hat r_{\m}\d n^\m +\hat r_L n^\m \d e^L_\m) dS.
\end{align}
As expected, this surface term alone fails to reproduce the surface term in \Ref{daje1}. This is fixed by the correcting term \Ref{defalpha}, which gives
\begin{align}\label{DeRossi}
& s\,d\a(\d)=-sP_{IJKL}\int_{\p\Si} e^I\w e^J e^{\r K}\d e^L_\r 
= s\int_{\p\Si} (n^\m \hat r_{I}\d e^I_\m - \hat r^\m n_{I}\d e^I_\m) dS 
+ \f{s}\g\int_{\p\Si}e^I\w\d e_I
\end{align}
The piece in $1/\g$ cancels the last term of the second row of \Ref{maggica1} when torsion vanishes, see \Ref{maggica3}.
Adding up \Ref{Dzeko} and the $\g$-less part of \Ref{DeRossi}, and using 
$\d(e^I_\m 
n^\m 
\hat r_I)=0$  we obtain
\be\label{Manolas}
-s(2\hat r_{\m}\d n^\m +\hat r_I n^\m \d e^I_\m+ \hat r^\m n_{I}\d e^I_\m) =
-s(2\hat r_{\m}\d n^\m -\hat r_\m \d n^\m -n^I\d \hat r_I - n_\m\d\hat r^\m -\hat r^I\d n_I)=
-s\hat r_{\m}\d n^\m,
\ee
where the final equality follows from $\hat r^I\d n_I=- n_I\d\hat r^I$ which cancels the third with the fifth term, and 
$n_\m\d\hat r^\m = -\hat r^\m\d n_\m 
= (s/2) \hat r^\m n_\m n_\r n_\s\d g^{\r\s}= 0$ which cancels the fourth. 
We have thus proved \Ref{daje}.

%----------------------------------------------------------------------------
\section{Poincar\'e charges at spatial infinity}\label{SecPoincare}
%----------------------------------------------------------------------------
Since the modification we propose changes the pre-symplectic form, we should check that it does not spoil established results, such as the recovery of Poincar\'e charges at spatial infinity 
with $\L=0$.
It was proved in \cite{Ashtekar:2008jw} that the original symplectic potential \Ref{Th0} vanishes on ${\cal T}_\infty$, a necessary condition for the canonical split without reducing the phase space, and that it leads to the correct Poincar\'e charges as in the metric formalism. %
Furthermore, the authors showed that the non-gauge-invariance of \Ref{Th0}  vanishes in the limit to $i^0$. This already signals that our modification will vanish in that limit, hence preserving those results. Let us show this explicitly, using the boundary and fall-off  conditions of  \cite{Ashtekar:2008jw}.\footnote{These we recall are slightly stronger than strictly necessary, as they are chosen also to eliminate the logarithm and supertranslation freedoms from the asymptotic symmetry group. It would be of course interesting to study relaxations admitting supertranslations, see e.g. \cite{Henneaux:2018cst}, as motivated by \cite{Hawking:2016sgy}.} 
 
 One chooses a reference flat metric $\eta_{\m\n}$ for the asymptotic behaviour, with hyperbolic slicing given by $\r^2:=\eta_{\m\n}x^\m x^\n$ and three angles collectively denoted by $\Phi$. Then the fall-off conditions appropriate to Poincar\'e symmetries are given for the tetrad by 
\be
e_\m^I= \,^{0}\!e_\m^I(\Phi) +\f{\,^{1}\!e_\m^I(\Phi)}\r+O(\r^{-2}),
\ee
with 
\be\label{e1}
\,^{0}\!e_\m^I(\Phi) = \d_\m^I, \qquad \,^{1}\!e_\m^I(\Phi) = \s(\Phi) (2\r_\m \r^I - \,^{o}\!e_\m^I),
\ee
$\s(\Phi)$ a reflection-symmetric arbitrary scalar function and 
$
\r_\m
:=\p_\m \r.$

We then have at leading order
\begin{align}\nn
& 
\int_{\p\Si}
\d_1\a(\d_2) - \d_2\a(\d_1) 
\\& \qquad =-P_{IJKL}\int_{\p\Si} \Big[ \Big(2\,^{0}\!e_\m^{[I} (\d_1 \,^{1}\!e_\n^{J]}) \,^{0}\!e^{\r K} - \,^{0}\!e_\m^{[I}  \,^{0}\!e_\n^{J]} ( \d_1\,^{1}\!e^{\r K}) \Big)(\d_2 \,^{1}\!e_\r^L)  - (\d_1\leftrightarrow \d_2) \Big] \f1{\r^{2}} dS^{\m\n}
\end{align}
which vanishes exactly using \Ref{e1} and the antisymmetry in $KL$.
The exact form we added gives no leading contribution to the pre-symplectic form in the limit to $i^0$, and the recovery of the Poincar\'e charges established in \cite{Ashtekar:2008jw} is left unaffected.

%%----------------------------------------------------------------------------
\section{Bifurcating horizons and the first law}\label{SecFirst}
%%----------------------------------------------------------------------------

We now show that our symplectic potential permits to derive the first law of black hole mechanics from the Noether charge associated with a Lie derivative, just like in the metric case \cite{Iyer:1994ys}. 
For the application of the formalism to derive the first law of black hole mechanics, we take a stationary and axisymmetric background solution, 
$\L=0$,
and choose $\Si$ to be a Cauchy hypersurface with two boundaries, one at the bifurcation surface $\cal B$ and one at spatial infinity $S_\infty$.
We take $\xi^\m$ to be the Killing vector that generates the horizon. Consider first the metric case. Since $\xi^\m$ is Killing, all variations $\d_\xi$ vanish and by linearity also the Hamiltonian charge,
\be\label{Ominv}
\dt H_\xi = \Om_{\scr EH}(\d,\d_\xi)=0. 
\ee 
Recalling the expression \Ref{defHxib} in terms of the Noether charge, we find a conservation law between surface charges at the bifurcating surface and at spatial infinity,
\be\label{firstlaw1}
\int_{\cal B} \d \k(\xi) = \int_{S_\infty} \d \k(\xi)- \th_{\scr EH}(\d)\lrcorner\xi,
\ee
where we used that fact that $\xi^\m|_{\cal B}=0$. If the perturbations are asymptotically flat and solution of the linearized field equations (but otherwise general), this equation evaluates to the first law of black hole mechanics (see e.g. \cite{Iyer:1994ys})\footnote{Notice that in this situation the (trivial) Hamiltonian charge is integrable, so both sides of \Ref{firstlaw1} are total variations: this is manifest for the LHS, for the RHS it follows from the standard ADM energy result plus the fact that $\p_\phi$ is tangent to $S_\infty$. This property is on the other hand not manifest in the final expression \Ref{firstlaw2} of the first law, where it is guaranteed by identities relating the variations of the various quantities appearing. }
\be\label{firstlaw2}
2k\d A = \d M - \Om_{H}\d J.
\ee
Crucially, this first law is invariant under $\th\mapsto\th+d\a$, since
the contribution of this ambiguity to \Ref{Ominv} always vanishes:
\be\label{ainv}
d \big( \d\a(\d_\xi) - \d_\xi\a(\d) \big) = 0. 
\ee
To see this, use the fact that $\a(\d)$ depends linearly on the variations and that $\d_\xi=0$ on the background fields. The quantity in square brackets then gives $\a(\d\d_\xi)-\a(\d_\xi\d)=0$ since $[\d_\xi,\d]=0$.

If the same state of affairs held in the tetrad formalism, we would agree with the argument given in \cite{TedMohd,Prabhu:2015vua}: neither options presented in \Ref{ThAch} give the Noether charge of the  metric theory, and since the first law should be invariant under redefinitions of the symplectic potential, we are left with the only possibility of looking for a new transformation to which the first law should be associated. 
The problem we see with this argument is the assumption that \Ref{Ominv} still holds in the tetrad theory, namely the requirement that for a Killing vector, $\pounds_\xi e^I=0$. This is not necessary, and can lead to inconsistencies; it is enough to require that 
\be\label{eKilling}
L_\xi e^I=\l_\xi{}^I{}_J e^J, 
\ee
since this automatically preserves the metric. Contracting with the inverse metric, we get an expression for the gauge transformation:
\be\label{deflxi}
\l_\xi{}^{IJ} = -e^{\r I}L_\xi e^{J}_\r = -D^{[I}\xi^{J]},
\ee
where we used \Ref{id4}
in the absence of torsion.\footnote{The reader may worry whether the invariance up to a gauge transformation of the tetrad under an isometry is consistent with the transformation of $\om^{IJ}$, namely whether
\[
L_\xi\om^{IJ}=F^{IJ}\lrcorner\xi \stackrel{?}{=}-d_\om\l^{IJ}_\xi=d_\om D^{[I}\xi^{J]}.
\]
The equality is indeed satisfied as it is nothing but the familiar Killing identity ${R_{\s\m\n\r}\xi^\s=\na_\m\na_\n\xi_\r}$ expressed in the tetrad formalism.
}
 Notice that it does not vanish on a bifurcating surface where $\xi^\m=0$. 
This immediately means that
\be\label{dHomom}
\dt H_\xi=\Om(\d,L_\xi)=\Om(\d,\d_{\l_\xi}),
\ee
namely the Killing diffeomorphism generator for a general potential is an internal Lorentz charge.

Using the gauge-invariant symplectic potential $\Th'_{\scr EC}$, the Lorentz charge is zero, see \Ref{due}, and thus from \Ref{dHomom} the vanishing  $\dt H_\xi=0$
of the diffeomorphism generator associated with a Killing vector is preserved. The Noether charge contains the exact Komar form, see \Ref{uno} (there is a priori another ambiguity in the cohomology of $\k_\xi$, but this is irrelevant for the first law since the boundary of a boundary is zero), and the symplectic potential reduces to the one of the Einstein-Hilbert action, see \Ref{daje}. 
Hence \Ref{dHomom} gives back precisely the conservation law \Ref{firstlaw1}, and the first law follows as usual. We conclude that our gauge-invariant potential associates naturally the first law to the invariance of the action under (covariant) Lie derivatives.

One may wonder whether the invariance of the first law under the ambiguity \Ref{ThAmbi} is lost. This is  not the case. In fact, we now show that the first law can also be derived from the non-gauge-invariant potential \Ref{Th0} and the same Lie derivative, without need for the non-linear object \Ref{defLieK} of \cite{TedMohd} or the automorphism construction of \cite{Prabhu:2015vua}, \emph{provided} one takes into account the presence of a non-zero Lorentz charge. 
Starting from the non-gauge-invariant potential \Ref{Th0} and using \Ref{eKilling}, the Hamiltonian generator \Ref{dHomom} for a Killing vector does not vanish anymore but coincides with the Lorentz generator.\footnote{A fact that can also be taken as motivation to prefer the gauge-invariant potential.} 
This evaluates to
\begin{align}\label{HxiL}
\Om_{\scr EC}(\d,L_\xi)=\Om_{\scr EC}(\d,\d_{\l_\xi}) = - \int_{\p\Si} P_{IJKL} \d(e^I\w e^J)\l_\xi{}^{KL},
\end{align}
where we used \Ref{Omegaded}, which is valid also for a field-dependent gauge parameter like $\l_\xi$ since the contribution from its variation is cancelled by the commutator term $\Th_{\scr EC}([\d,\d_{\l_\xi}])$ which is in this case not vanishing.

To evaluate the hand side, we could compute the Noether current associated with $L_\xi$, but we can also use \Ref{defL} and the bilinearity of the symplectic form to derive
\begin{align}\label{Oml}
\Om_{\scr EC}(\d,L_\xi) &= \Om_{\scr EC}(\d,\pounds_\xi) +\Om_{\scr EC}(\d,\d_{\om\lrcorner\xi}). 
\end{align}
The first piece gives
\begin{align}
\Om_{\scr EC}(\d,\pounds_\xi) &= \int_{\p\Si} \d j(\pounds_\xi) -\th_{\scr EC}(\d)\lrcorner\xi
=\int_{\p\Si} P_{IJKL} \d(e^I\w e^J \, \om^{KL}\lrcorner\xi) -\th_{\scr EH}(\d)\lrcorner\xi +d\a(\d)\lrcorner\xi \nn\\\label{heid}
& =\int_{\p\Si} P_{IJKL}\Big[e^I\w e^J\, \d \l_\xi^{KL} + \d(e^I\w e^J) \, \om^{KL}\lrcorner\xi \Big] -\th_{\scr EH}(\d)\lrcorner\xi,
\end{align}
where in the second equality we used  \Ref{Ach} on-shell, and the equivalence $(\th_{\scr EC}+d\a)|_{\om^{IJ}=\og{\om}^{IJ}} =\th_{\scr EH}$ previously established; in the last step we used
\be
\int_{\p\Si} d\a(\d)\lrcorner\xi =\int_{\p\Si} \pounds_\xi\a(\d)= \int_{\p\Si} P_{IJKL}\Big[ e^I\w e^J\, \d \l_\xi^{KL} - e^I\w e^J \,\d \om^{KL}\lrcorner\xi \Big]. 
\ee
This can be proved by explicit calculation using \Ref{defalpha} and \Ref{deflxi}, but also observing that  
\be
\pounds_\xi\a(\d) = L_\xi\a(\d) = \d_{\l_\xi}\a(\d) +\a([\d,\d_{\l_\xi}-\d_{\om\lrcorner\xi}])=\a([\d,\d_{\l_\xi}-\d_{\om\lrcorner\xi}]),
\ee
which follows using $\d_\l\a(\d)=0$ for a gauge transformation and \Ref{deflxi} for the background fields $\phi$:
\be
L_\xi\a(\phi,\d\phi) = \a(L_\xi\phi,\d\phi)+\a(\phi,L_\xi\d\phi) = \d_{\l_\xi}\a(\phi,\d\phi)  -\a(\phi,[\d_{\l_\xi},\d]\phi) + \a(\phi,[L_\xi,\d]\phi)
\ee
and $[L_\xi,\d]=[\d_{\om\lrcorner\xi},\d]$.
The second piece in \Ref{Oml} is again a Lorentz charge, 
\be\label{LC2}
\Om_{\scr EC}(\d,\d_{\l_\xi}) = - \int_{\p\Si} P_{IJKL} \d(e^I\w e^J)\, \om^{KL}\,\lrcorner \xi,
\ee
and cancels the second term in \Ref{heid}.
We can now equate \Ref{HxiL} to \Ref{Oml} with the above manipulation, and derive
\be\label{firstlaw3}
 - \int_{\p\Si} P_{IJKL} \d(e^I\w e^J)\l_\xi{}^{KL} = \int_{\p\Si} P_{IJKL}e^I\w e^J\, \d \l_\xi^{KL} -\th_{\scr EH}(\d)\lrcorner\xi.
\ee
Finally, notice that
\be
\int_{\p\Si}  P_{IJKL} \d(e^I\w e^J\, \l_\xi^{KL})=- \int_{\p\Si}  P_{IJKL} \d(e^I\w e^J D^{K}\xi^{L}) = \int_{\p\Si}\d \k_\xi,
\ee
hence \Ref{firstlaw3} gives the same identity \Ref{firstlaw1} as the metric and gauge-invariant symplectic potential calculations, from which the first law follows.
The Lorentz charge is thus crucial to recover the Komar form and the first law using the non-gauge-invariant symplectic potential and the ordinary Lie derivative.

We conclude that also with the original potential \Ref{Th0} the first law follows from the Noether identity and Lie derivatives. This is consistent with the findings of \cite{Ashtekar:2000hw}, where the first law for stationary black holes and more in general for isolated horizons was recovered from the second equality in \Ref{defHxia}, without going through the Noether current expression \Ref{defHxib}. In \cite{Ashtekar:2000hw} the internal Lorentz symmetry at the isolated horizon was fixed, and the non-vanishing of \Ref{dHomom} indeed noticed, and referred to as  horizon energy. Our results show that this is nothing but the Lorentz charge.

The bottom line of the derivation of the first law \Ref{firstlaw3} with the non-gauge-invariant potential and the Lie derivative is that the Komar charge, absent from the symplectic potential, pops up through the Lorentz charge giving the diffeomorphism Hamiltonian generator. This simple reshuffling restoring the first law extends to any symplectic potential in the equivalence class \Ref{ThAmbi}.
Therefore, there still is a perfect invariance of the first law under the cohomology ambiguity in the symplectic potential, albeit in a subtler way than in the metric case. The subtlety is that adding an exact form to the symplectic potential can introduce surface Lorentz charges, which in turn provide non-zero charges also for the Hamiltonian generators of Killing isometries. These have to be taken into account if one wants to recover the first law from the covariant Lie derivative alone.

Let us also compare our results with those of \cite{TedMohd,Prabhu:2015vua}. There it was acknowledged that the symplectic potential is not gauge-invariant, and it was shown that one can still work with it and define Hamiltonian diffeomorphism charges vanishing for Killing vectors, provided these diffeomorphism are not associated with Lie derivatives alone, standard or covariant, but with automorphisms of the tetrad. 
This construction uses the non-linear object \Ref{defLieK} whose action depends on the tetrad also when acting on other fields, and whose extension in the case of an affine connection with torsion is unclear to us. Our findings show that there is a simpler alternative: keep the covariant Lie derivative and switch to a gauge-invariant potential, or keep the non-gauge-invariant potential but take into account the Lorentz charges and the non-vanishing of \Ref{dHomom}. This said about our alternative, we remark that the motivations of \cite{TedMohd,Prabhu:2015vua} include topological issues and smoothness of fields; we have not looked at these aspects, so we are not in a position to assess how they would change our results.

%%----------------------------------------------------------------------------
\section{Conclusions}
%%----------------------------------------------------------------------------

In this paper we have proposed a gauge-invariant symplectic potential for tetrad general relativity,
implementing what discussed for generic gauge theories in \cite{Barnich:2007bf}. See also \cite{DonnellyFreidel,Gomes:2018shn} for additional discussions on the importance of gauge-invariance of the phase space.
Our construction uses the freedom to add exact spacetime forms, namely corner terms, to the symplectic potential. A gauge-invariant symplectic potential cannot be directly read off from the action, but additional input is required in the choice of the right corner term, which in turns determines the covariant phase space and resulting Hamiltonian fluxes/surface charges.
The gauge-invariant potential eliminates what we see as spurious internal Lorentz charges produced by the symplectic potentials used so far in the literature. It does not change the Poincar\'e charges at spatial infinity, since the gauge-breaking terms vanish in that limit. It plays a key role on the other hand in deriving the first law of black hole mechanics from the Noether identity associated with the Lie derivative and a vanishing Killing Hamiltonian flux, like in the metric theory. 

We also pointed out that the derivation of the first law from the covariant Lie derivative is in the end invariant under the cohomology ambiguity in the symplectic potential, and thus independent of having chosen a gauge-invariant one: it suffices to take into account the non-trivial Lorentz charges that can be present.
The technical statement is that the invariance of the first law under the ambiguity $\th\mapsto\th+d\a$, which is guaranteed in the metric theory by the fact that for a Killing field $\Om(\d,\d_\xi)=0$, is now guaranteed by the $\Om(\d,L_\xi)=\Om(\d,\d_{\l_\xi})$, and therefore the first law is recovered with a 
\emph{non-vanishing} Killing Hamiltonian flux, if one uses the original potential, and a vanishing Killing Hamiltonian flux if one uses the gauge-invariant symplectic potential.

Our gauge-invariant symplectic potential turns out to be exactly equivalent to the Einstein-Hilbert symplectic potential when torsion vanishes, for arbitrary variations. This was not granted a priori since they could have differed by gauge-invariant exact 2-forms, e.g. terms written directly as variation of the metric. The proof was based on some identities for differential geometry with tetrads that allows us to recover variations of extrinsic curvature and 2d corner terms.

For simplicity, we have neglected in this paper boundary terms in the action, and the explicit contribution of matter fields. Boundary terms and topological terms can be added following the previous treatments \cite{Corichi:2016zac,SorkinCorner17}. The matter contribution is worth exploring: Having settled the issue of the equivalence of the symplectic potential when torsion vanishes, this can be used to study the contribution of torsion to the charges.

Among the applications of our results we mention
the study of boundary degrees of freedom, in particular the $1/\g$ term in \Ref{RB30} has been shown to lead to an interesting description in terms of a conformal field theory on the boundary \cite{Freidel:2016bxd} (see also \cite{Wieland:2017cmf,Geiller:2017whh}) and it would be interesting to see if and how that description is affected by our results. A related issue concerns calculations of entanglement entropy with the action \Ref{SEC}, see e.g. \cite{Ashtekar:2004cn,Bodendorfer:2014fua}.
Finally, approaches to quantization suggest to endow the covariant phase space methods within the Batalin-Fradkin-Vilkovisky framework, which for the non-gauge-invariant potential \Ref{Th0} has been discussed in \cite{Cattaneo:2017kkv}.

Throughout the paper we restricted attention to non-null hypersurfaces. 
Quasi-local charges and conservation laws are even more interesting when one considers null hypersurfaces (see e.g. \cite{Ashtekar:1981bq, Wald:1999wa,MikeNull,Hawking:2016sgy,Wieland:2017zkf,Hopfmuller:2018fni}), and it is natural to ask how our results extend to that case. A symplectic potential for tetrad gravity giving vanishing internal Lorentz charges can also be obtained when the 2d corner hinges between a space-like and a null hypersurface \cite{Wieland:2017zkf,AbhayWolfgang}.
We explored the Hamiltonian structure of Einstein-Cartan gravity on null hypersurfaces in \cite{IoSergeyNull,IoElena}, and we plan in future work to look at the covariant phase space perspective on them.

%--------------------------------------------------------------------------------------------------
%--------------------------------------------------------------------------------------------------
\subsection*{Acknowledgments}
%--------------------------------------------------------------------------------------------------
%--------------------------------------------------------------------------------------------------

\appendix
We are grateful to Abhay Ashtekar and Wolfgang Wieland for discussions on covariant phase space methods and observables. We thank Norbert Bodendorfer and Ted Jacobson for feedback and discussions.

%----------------------------------------------------------------------------
\section*{Appendix}
%----------------------------------------------------------------------------
%----------------------------------------------------------------------------
\setcounter{equation}{0}
\renewcommand{\theequation}{A.\arabic{equation}}

We define the spacetime Levi-Civita density $\ut{\eps}_{\m\n\r\s}$ as the completely antisymmetric object with $\ut{\eps}_{0123}=1$, and 
$\tl\eps^{\m\n\r\s}\ut{\eps}_{\m\n\r\s}=-4!$.
We denote the spacetime volume form as
\be
\eps:=\f1{4!}\eps_{\m\n\r\s}dx^\m\w dx^\n\w dx^\r\w dx^\s, \qquad \eps_{\m\n\r\s}:=\sqrt{-g} \, \ut{\eps}_{\m\n\r\s}.
\ee
The Hodge dual $\star: \L^p\mapsto\L^{n-p}$
is defined in components as
\be
(\star \om^{(p)})_{\m_1..\m_{n-p}} := \f{1}{p!} \om^{(p)}{}^{\a_1..\a_p} \eps_{\a_1..\a_p\m_{1}..\m_{n-p}}.
\ee

For (non-null) hypersurfaces, we use the following conventions: if the Cartesian equation of $\Si$ is $\varphi(x)=0$, the unit normal is
\be
n_\m:= \f{s}{\sqrt{g^{\r\s}\p_\r\varphi\p_\s\varphi}} \p_\m \varphi, \qquad s:=n^2=\pm1,
\ee
and the induced volume form
\be
\eps^\Si:=\eps\lrcorner n, \qquad \eps^\Si_{\m\n\r}:=n^\s\eps_{\s\m\n\r}, \qquad
d\Si_\m = sn_\m d\Si.
\ee
On a space-like surface $S$ within $\Si$, with unit normal $\hat r_\m$, we have $\hat r^2=-s$ and the induced area form
\be
\eps^{S}:=\eps^{\Si}\lrcorner \hat r, \qquad \eps^{S}_{\m\n}:=n^\r\hat r^\s\eps_{\m\n\r\s}.
\ee

For the internal Levi-Civita density $\eps_{IJKL}$ we refrain from adding the tilde. We keep the same convention, ${\eps}_{0123}=1$, hence the tetrad determinant is
\be\label{tetId2}
e = -\f1{4!}\eps_{IJKL}\tl\eps^{\m\n\r\s} e_\m^I e_\n^J e_\r^K e_\s^L, \qquad 4ee^{[\m}_Ie^{\n]}_J = - \eps_{IJKL}\tl\eps^{\m\n\r\s}e^K_\r e^L_\s,
\ee
and we take $e>0$ for a right-handed tetrad.

\providecommand{\href}[2]{#2}\begingroup\raggedright\endgroup

%--------------------------------------------------------------------------------------------------
%--------------------------------------------------------------------------------------------------
\end{document}